\documentclass[runningheads]{llncs}
\usepackage[T1]{fontenc}
\usepackage{newtxtext}       %
\usepackage[varvw]{newtxmath}       
\usepackage{titlesec}

\titlespacing{\paragraph}{%
  0pt}{
  0.5\baselineskip}{
  1em}%
\usepackage{graphicx}
\usepackage{booktabs}

\usepackage{hyperref}
\usepackage{subcaption}
\usepackage{cleveref}
\usepackage{color}
\usepackage{booktabs}
\usepackage{multirow}
\usepackage{array}

\newcommand{\inlineemoji}[1]{\includegraphics[height=1em]{emojis/#1.png}}
\usepackage{fontawesome5}

\newcommand{\hf}{\inlineemoji{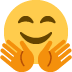}}

\newcommand{\vmark}{\inlineemoji{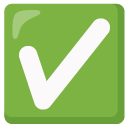}}
\newcommand{\xmark}{\inlineemoji{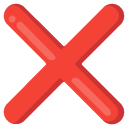}}
\newcommand{\githubrepo}[1]{%
    \begin{center}
        \href{#1}{\small \faGithub \ \texttt{#1}}
    \end{center}
}

\newcommand{\micep}{MICE-P}

\begin{document}

\title{MICE to Meet You: \\From Re-ranking Only to End-to-End Retrieval}

\titlerunning{MICE to Meet You: From Re-ranking Only to End-to-End Retrieval}

\author{Victor Morand \inst{1}  \and Yuxuan Zong\inst{1} \and Mathias Vast\inst{1,2} \and Josiane Mothe\inst{3} \and Benjamin Piwowarski \inst{2} }
\authorrunning{V. Morand et al.}

\institute{Sorbonne Universit\'e, CNRS, ISIR, Paris, France \and Sinequa by ChapsVision, Paris, France \and Univ. Toulouse, IRIT, CLLE, CNRS, Toulouse, France\\ \email{\{name\}@isir.upmc.fr}}
\maketitle              
\begin{abstract}

Late-interaction models offer state-of-the art efficiency-effectiveness trade-offs for neural retrieval by enabling document representations to be pre-computed and efficiently indexed. Yet, their relatively simple interaction mechanisms, typically MaxSim, can limit their effectiveness compared to cross-encoders.
MICE, a recently introduced re-ranking architecture,
aims to combine cross-encoder-like effectiveness through expressive transformer-based scoring with reduced inference-time computation through document-side pre-computation. 
However, no practical way to store these document representations was explored. 
We address this limitation here and introduce \micep{}, the first end-to-end retrieval pipeline with MICE, based on PLAID to index the contextualized document token representations.
We conduct a controlled comparison of MICE, ColBERT and cross-encoders by training comparable models using the Ettin Family 
under matched experimental conditions.
Our results show that \micep{} defines a new class of retrieval architecture with an efficiency–effectiveness trade-off of its own: it matches late-interaction retrieval throughput while retaining the effectiveness of cross-encoders. 

\end{abstract}

\githubrepo{https://github.com/xpmir/mice}

\section{Introduction}

Multi-stage ranking pipelines balance retrieval efficiency and effectiveness by combining an \textit{efficient} first-stage retriever (e.g., BM25 \cite{BM25}, bi-encoder \cite{karpukhin-etal-2020-dense}) with a more \textit{effective} but less efficient re-ranker (e.g. cross-encoder \cite{monobert}). 
Recent neural retrieval models offer more favorable efficiency--effectiveness trade-offs. Late-interaction models such as ColBERT \cite{khattab2020colbert} pre-compute document token representations and rely on specialized indexes such as PLAID \cite{santhanam2022plaid} for efficient (but space--intensive) storage, candidate generation, and retrieval. This  makes them well-suited as end-to-end retrieval systems. 
However, they still tend to be less effective than cross-encoders \cite{rosa2022defensecrossencoderszeroshotretrieval}.

\noindent This leaves a gap between the expressive but costly scoring of cross-encoders and the efficient but more constrained interactions of late-interaction models. 
Several architectures have explored intermediate designs between independently encoded retrieval models and fully joint cross-encoders. Mid-fusion approaches such as PreTTR \cite{macavaneyEfficientDocumentReRanking2020} and MORES \cite{gaoModularizedTransfomerbasedRanking2020mores} process queries and documents separately in early layers and jointly in later ones. Starting from a cross-encoder, MICE \cite{emnlpmice} follows this direction and aims to improve efficiency while largely retaining effectiveness by (1)~removing superfluous interactions from the costly self-attention mechanism \cite{attention}, and (2)~enabling the pre-computation of document representations. It is composed of two parts: the \textit{contextualization layers}, where document and query representations are contextualized independently, and the \textit{interaction layers}, where frozen document representations interact with the query to detect matching signals. By doing so, it preserves richer interactions than MaxSim while reducing inference-time computation further than mid-fusion approaches by limiting those interactions. 
More recently, the ``Fast But Not Late'' KaLM-Reranker \cite{zhaoKaLMRerankerV1FastNot2026} similarly pre-computes document-side representations before introducing query--document interactions at scoring time, but within an encoder-decoder architecture \cite{zhang2025t5gemma2seeingreading}, thus less suited for the IR task than an encoder-only \cite{ettin}. 

MICE's pre-computed document representations make indexed retrieval possible in principle. The original work~\cite{emnlpmice}, however, only studies re-ranking, and analyze efficiency only theoretically, leaving open how to store and search its fine-grained representations over a full collection.
These challenges have been explored for late-interaction systems, through vector compression as in ColBERTv2 \cite{santhanam2022colbertv2}, or product quantization \cite{PQ}. PLAID \cite{santhanam2022plaid} combines compressed multi-vector storage with centroid-based candidate generation and pruning. As its retrieval procedure is tightly coupled to MaxSim, it is unclear whether the same indexing can be transferred to MICE.

We explore this issue and introduce \micep{} by adapting MICE to use PLAID, enabling efficient storage of its fine-grained document representations.
We also revisit MICE training with recent training strategies \cite{aarsen2026ettin-reranker} 
and successfully learn ModernBERT-based MICE models. 
Together, these contributions establish \micep{} as a new class of retrieval architecture that offers a more favorable efficiency–effectiveness trade-off than late-interaction scorers.
Specifically, we ask the following questions:
\begin{enumerate}
    \item[\textbf{RQ1}] Can we build an efficient MICE-based end-to-end retriever?
    \item[\textbf{RQ2}] Can MICE benefit from ModernBERT and match the effectiveness of comparable cross-encoder and late-interaction models?
\end{enumerate}

\noindent Our code, configurations and all checkpoints are released at \href{https://github.com/xpmir/mice}{\small \faGithub \ \texttt{github.com/xpmir/mice}}.


\section{\micep(LAID)}

\micep{} addresses two limitations of MICE  \cite{emnlpmice}. 
First, we turn theoretical gains from document pre-computation into reality by introducing a low-dimensional bottleneck (similar to PreTTR \cite{macavaneyEfficientDocumentReRanking2020}) that maps MICE's contextualized document representations to a representational space compatible with PLAID. We show that this does not impact the effectiveness of the interaction layers. 
Second, we develop a transfer strategy that starts from an already trained cross-encoder and progressively adapts it to the MICE architecture. This allows ModernBERT-based MICE models to retain the effectiveness of their cross-encoder counterparts.

\subsection{From MICE to \micep: Using PLAID to Store Documents}\label{mice_spec}

Even if PLAID \cite{santhanam2022plaid} targets MaxSim-based models such as ColBERT \cite{khattab2020colbert}, not attention-based interaction as in MICE \cite{emnlpmice}, we can use its residual format for compressing and storing documents representations. 
To limit the size of the vectors to store, we insert a bottleneck on the contextualized document representations: a linear projection to dimension 128 (matching LateOn dimension for a fair comparison as \texttt{Ettin-68m} uses a hidden size of 512 and k-means centroids degrade in higher dimensions), followed by LayerNorm \cite{ba2016layernormalization} for normalization (FastPLAID \cite{fastplaid2025} indexes normalized vectors). The indexed 128-d vectors are then decompressed and a linear up-projection restores the original model dimension for the interaction layers, without significative drop in effectiveness (58.0 nDCG@10 points on average on BEIR13 for MICE-61m vs 57.8 for MICE-P-61m, from Table~\ref{tab:full-results}).
While representations are stored using the ColBERTv2 residual format (via FastPLAID), we bypass PLAID's inverted-index retrieval pipeline and only decompress candidate representations for the interaction layers. \micep{} thus still relies on a distinct first-stage retriever, DistilBERT-SPLADEv3 \cite{lassance2024spladev3}, to retrieve the top passages for re-ranking.

\subsection{Transferring Cross-Encoders to MICE}\label{transfer_strat}

\noindent \textbf{ Fine-tuning.} MICE was originally initialized from a pre-trained encoder and distilled with Margin-MSE \cite{emnlpmice}.
We take here a different approach: we start from an existing cross-encoder
(\texttt{Ettin-reranker} checkpoint) and fine-tune it in two phases: (1) we adapt the cross-encoder weights to the new architecture by continuing training on the ranking task; (2) we then add the low-dimensional bottleneck (Section~\ref{mice_spec}) and train it together with the interaction layers, keeping contextualization layers frozen, since back-propagating through the fresh bottleneck degraded them.
For fair comparisons, we use the same protocol to fine-tune ColBERT counterparts; the released \texttt{Ettin-reranker} serve as comparisons for cross-encoders.
We follow \texttt{Ettin-reranker} training strategy \cite{aarsen2026ettin-reranker}and apply a pointwise MSE loss on a large distillation dataset\footnote{\href{https://huggingface.co/datasets/cross-encoder/ettin-reranker-v1-data}{\texttt{cross-encoder/ettin-reranker-v1-data}}.} : 143M query--passage pairs (e.g., MS~MARCO or TriviaQA) scored by the \texttt{mxbai-rerank-large-v2} cross-encoder, and drawn from the same public collections as LateOn \cite{sourty2026denseonlateonfullyopen}, making it possible to compare all architectures on common ground.
Differently, we finetune ColBERT variants with a Pearson loss \cite{pearsonVIIMathematicalContributions1896}, close to MSE and accounting for the cross-encoder teacher's scale does not matching late-interaction score distribution.

\noindent \textbf{Validating the training strategy.} We train two reference models:  
(1) a cross-encoder based on the \texttt{Ettin-68m} backbone and
(2) a ColBERT model from  unsupervised LateOn checkpoint\footnote{\texttt{lightonai/LateOn-unsupervised}}.
Table~\ref{tab:full-results} confirms that (1) our reproduced \texttt{Ettin-reranker-68m} matches the released checkpoint \cite{aarsen2026ettin-reranker} (58.0 vs. 58.1 average nDCG@10 on both BEIR13 and LoTTE) and (2) \texttt{LateOn-Pearson} is close to the released LateOn \cite{sourty2026denseonlateonfullyopen} on BEIR13 (57.2 vs. 58.6, without PLAID storage), despite the latter additionally using hard negatives and InfoNCE \cite{oord2019representationlearningcontrastivepredictive}, validating our training strategy for ColBERT models.

\section{Experimental Setup}

\begin{figure}[t]
    \centering
    \vspace{-0.2cm}
    \includegraphics[width=0.9\linewidth]{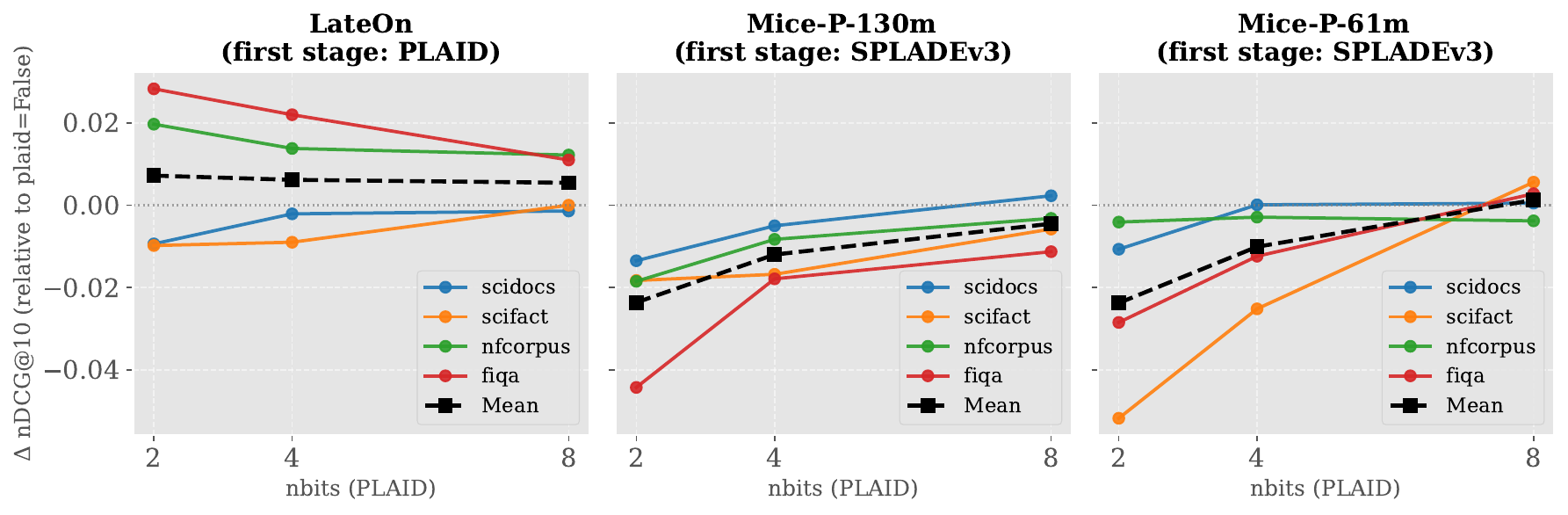}
    \caption{Comparing the impact of quantization (\texttt{nbits}) on ranking effectiveness on five BEIR-decontaminated datasets, for \micep{} and ColBERT models at similar sizes. } \label{fig:abl_nbits}
    \vspace{-0.4cm}
\end{figure}

\noindent \textbf{Models.} We use \texttt{Ettin-68m} \cite{ettin} to determine the key design choices, and later extend them on \texttt{Ettin-150m} for comparing to the state-of-the-art \texttt{LateOn} (149M)  \cite{sourty2026denseonlateonfullyopen}.
Note that we kept the number of contextualization  and interaction layers in \micep{} consistent with the original MICE models \cite{emnlpmice} for each backbone.

\noindent \textbf{Tokenization.} Although ModernBERT \cite{modernbert} supports up to 8K tokens, we limit queries and documents to 32 and 512 tokens respectively. This keeps the comparison fair across architectures, in particular for ColBERT variants who are sensitive to query and document length due to MaxSim, and limits the size of PLAID document stores.

\noindent \textbf{Evaluation.} Following \cite{emnlpmice,lassance2024spladev3}, we evaluate on the 13 datasets publicly available from the decontaminated BEIR \cite{thakur_beir_2021} benchmark (BEIR13) of Sourty et al. \cite{sourty2026denseonlateonfullyopen}, since the official version overlaps with our training data, and validate on Nano-BEIR \cite{NanoBEIRMultilingualInformation2026}. 
For out-of-domain, we also evaluate our models over the Search version of the LoTTE benchmark \cite{santhanam2022colbertv2}, as these datasets are commonly used as reference when evaluating late-interaction models. 
We always re-rank the top-100 of DistilBERT-SPLADEv3 \cite{lassance2024spladev3} when using FastPLAID \cite{fastplaid2025} as a document store for \micep{} but not for ColBERT variants that can leverage PLAID as a first stage. Other hyperparameters are detailed in the repository.

\noindent \textbf{PLAID Configuration.} We index document representations with FastPLAID \cite{fastplaid2025}. The main parameter is the number of bits \texttt{nbits} used to quantize residuals, which trades index size for fidelity. We select it on five datasets from BEIR decontaminated \cite{sourty2026denseonlateonfullyopen}:SciDocs \cite{specter2020cohan}, SciFact \cite{Wadden2020FactOF}, ArguAna \cite{wachsmuth-etal-2018-retrieval}, NFCorpus \cite{nfcorpus}, and FiQA-2018 \cite{fiqa}, and report nDCG@10 drop relative to a baseline without compression (\Cref{fig:abl_nbits}). 
\texttt{LateOn} is insensitive to quantization down to 2 bits (mean $\Delta\approx-0.002$), consistent with the standard ColBERT setting \cite{santhanam2022colbertv2}, so we set \texttt{nbits}=2. \micep{} degrades sharply at 2 bits (mean $\Delta<-0.03$, up to $-0.05$ on FiQA and ArguAna) but stays close to 8 bits at 4 bits, which halves the residual storage. We set \texttt{nbits}=4. 

\section{Results}

\Cref{tab:full-results} details the full results for two sizes of Ettin encoders (68m and 150m) across BEIR13 and the average results on LoTTE-Search \cite{khattab2020colbert}.

\begin{table}[t]
\caption{
Detailed results on decontaminated BEIR-13 and average results on LoTTE-Search, in nDCG@10. ``Store'': indexed with PLAID; ``1st St'': PLAID (P) or SPLADE-v3 (S); ``Dim.'': document representation dimension. \hf~marks public checkpoints. \textbf{Bold} and \underline{underlined} respectively indicate best and second best model per size. }
\label{tab:full-results}
\resizebox{\textwidth}{!}{
\begin{tabular}{lccc|ccccccccccccc|cc}
\toprule
Scorer & Dim. & Store & 1st St & Ar & CF & DB & FE & Fi & HPQ & NFC & NQ & Q & SD & SF & T-C & T-v2 & BEIR13 & LoTTE-S \\
\midrule
\midrule
\hf \ Ettin-reranker-68m-v1 
& -- & \xmark & S & 35.8 & 44.0 & 33.2 & 92.8 & 51.0 & 79.6 & 23.8 & 96.2 & 91.3 & 6.8 & 87.6 & 78.5 & 34.4 & \textbf{58.1} & \textbf{58.1} \\
    \quad Our reproduction & -- & \xmark & S & 35.2 & 43.7 & 32.6 & 92.6 & 50.2 & 79.4 & 24.0 & 97.5 & 91.5 & 7.6 & 87.5 & 78.1 & 34.0 & \underline{58.0} & \underline{58.0} \\
    \cmidrule(lr){1-19}
    \quad MICE-61m & 512 & \xmark & S & 33.5 & 43.3 & 33.4 & 92.4 & 51.7 & 79.1 & 24.4 & 96.2 & 91.3 & 9.1 & 87.6 & 76.5 & 34.9 & \underline{58.0} & 57.3 \\
    \quad MICE-P-61m & 128 & \xmark & S & 33.5 & 43.4 & 33.6 & 92.2 & 52.3 & 78.8 & 24.8 & 94.3 & 91.2 & 9.0 & 86.6 & 77.2 & 34.6 & 57.8 & 56.9 \\
    \quad MICE-P-61m & 128 & \vmark & S & 32.6 & 42.9 & 33.1 & 92.2 & 51.1 & 78.3 & 24.5 & 95.3 & 91.1 & 9.0 & 84.0 & 75.8 & 34.7 & 57.3 & 56.0 \\
    \cmidrule(lr){1-19}
    \quad ColBERT-Ettin-68m & 128 & \xmark & S & 30.1 & 41.6 & 31.0 & 89.8 & 53.2 & 76.8 & 24.9 & 93.6 & 91.0 & 8.8 & 86.6 & 75.5 & 34.7 & 56.7 & 56.9 \\
    \quad ColBERT-Ettin-68m & 128 & \vmark & P & 26.0 & 29.9 & 21.7 & 19.6 & 49.6 & 47.9 & 26.0 & 65.1 & 89.9 & 6.8 & 81.8 & 71.6 & 31.0 & 43.6 & 54.3 \\
\midrule
\hf \ lightonai/LateOn (150m) & 128 & \xmark & S & 34.9 & 38.6 & 31.5 & 91.8 & 54.8 & 76.4 & 26.2 & 93.6 & 91.6 & 14.9 & 89.8 & 81.7 & 36.1 & \underline{58.6} & \textbf{61.3} \\
\hf \ lightonai/LateOn (150m) & 128 & \vmark & P & 33.6 & 38.2 & 28.1 & 90.9 & 57.6 & 71.8 & 28.2 & 87.9 & 89.1 & 14.0 & 88.8 & 79.0 & 33.1 & 57.0 & 58.2 \\
    \quad LateOn-Pearson (150m) & 128 & \xmark & S & 26.5 & 42.4 & 30.9 & 90.1 & 53.8 & 78.1 & 24.8 & 94.8 & 91.7 & 10.8 & 89.0 & 75.6 & 34.8 & 57.2 & 58.9 \\
    \quad LateOn-Pearson (150m) & 128 & \vmark & P & 24.3 & 35.8 & 25.7 & 29.9 & 51.9 & 63.3 & 24.9 & 46.2 & 91.0 & 8.8 & 87.8 & 73.9 & 32.4 & 45.8 & 57.3 \\
\midrule
\hf \ Ettin-reranker-150m-v1 & -- & \xmark & S & 37.2 & 44.7 & 33.9 & 92.8 & 50.7 & 79.7 & 23.9 & 97.6 & 91.9 & 9.8 & 88.5 & 79.1 & 34.3 & \textbf{58.8} & \underline{59.1} \\
    \quad MICE-P-130m & 128 & \xmark & S & 33.4 & 43.3 & 34.4 & 92.7 & 54.1 & 79.2 & 24.9 & 97.6 & 91.6 & 9.4 & 85.6 & 78.7 & 35.2 & 58.5 & 58.5 \\
    \quad MICE-P-130m & 128 & \vmark & S & 32.7 & 43.2 & 33.9 & 92.3 & 52.4 & 78.6 & 24.0 & 97.6 & 91.2 & 8.9 & 84.0 & 78.5 & 35.3 & 57.9 & 57.4 \\
    \cmidrule(lr){1-19}
    \quad ColBERT-Ettin-150m & 128 & \xmark & S & 33.4 & 42.7 & 32.0 & 89.6 & 51.5 & 77.4 & 24.8 & 94.7 & 91.1 & 10.7 & 86.6 & 77.2 & 34.9 & 57.4 & 57.4 \\
    \quad ColBERT-Ettin-150m & 128 & \vmark & P & 30.1 & 28.9 & 24.6 & 25.5 & 49.2 & 60.7 & 24.2 & 81.8 & 89.8 & 8.1 & 85.0 & 74.8 & 32.7 & 47.3 & 55.3 \\
\bottomrule
\end{tabular}
}
\end{table}

\noindent \textbf{Baselines.} First, our training strategy yields MICE models based on ModernBERT on par with their cross-encoder counterparts ($\Delta < 0.3$ and $\Delta < 1$ nDCG@10 on average respectively for BEIR13 and LoTTE-Search, without PLAID storage). Furthermore, all MICE models outperform their ColBERT counterparts by more than 1 nDCG@10 points on average on BEIR13, and matches or beats them on LoTTE-Search.
Together, these results confirm that ModernBERT can effectively be used as a MICE backbone (positively answering \textbf{RQ2}).  
Second, transferring MICE into \micep{} by applying a bottleneck and using PLAID to store the document contextualized representations, only marginally impacts the effectiveness on BEIR13 ($\Delta < 0.7$). On LoTTE-Search, \micep{} appears to be more impacted (around 1 nDCG@10 points).
Interestingly, our ColBERT checkpoints, including the reproduced \texttt{LateOn-Pearson}, collapse dramatically with PLAID as first-stage on BEIR13. The original \texttt{LateOn} maintains a more consistent effectiveness ($\Delta <1.6$ nDCG@10 points).

\begin{figure}[t]
    \centering
    \begin{subfigure}[b]{0.72\textwidth}
        \centering
        \includegraphics[width=\linewidth]{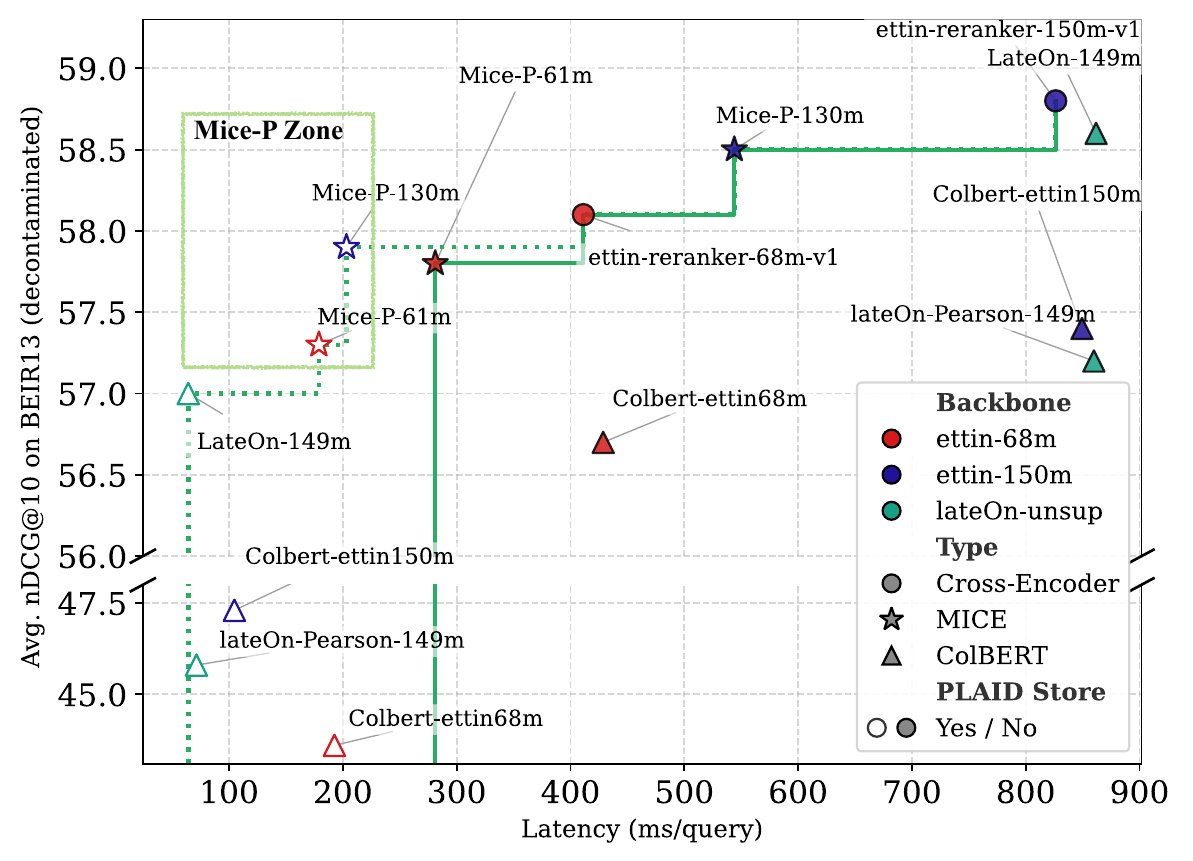}
        \caption{Pareto frontier for the studied retrieval pipelines.}
        \label{fig:pareto_compute}
    \end{subfigure}
    \hfill
    \begin{subfigure}[b]{0.23\textwidth}
        \centering
        \includegraphics[width=\linewidth]{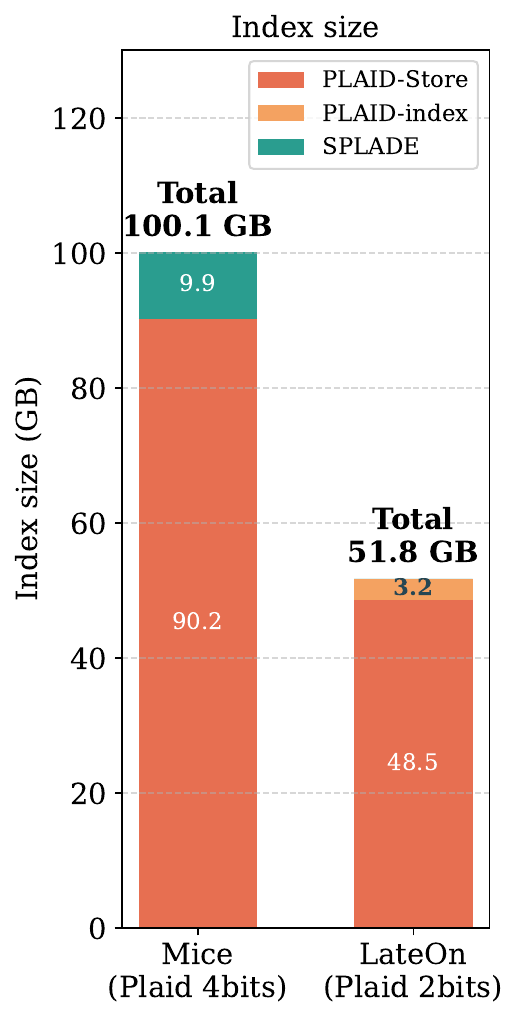}
        \caption{Index sizes on Climate-FEVER (5M docs)}
        \label{fig:pareto_mem}
    \end{subfigure}
    \caption{Comparing retrieval architectures across efficiency, effectiveness and storage.}
    \label{fig:pareto_fig}
    \vspace{-.3cm}
\end{figure}

\noindent \textbf{Mice-P is Pareto Optimal.} To answer more precisely to \textbf{RQ1}, we also measure the end-to-end retrieval latency of the different pipelines\footnote{We use Seismic \cite{bruch2024seismic} for fastest SPLADE retrieval, while cross-encoders and ColBERT models inference is based on sentence-transformers \cite{reimers-2019-sentence-bert} and FastPLAID \cite{fastplaid2025} respectively.} and of the different baselines, including ColBERT variants used as re-rankers to isolate the cost of the scoring from first-stage retrieval. \Cref{fig:pareto_compute} details the trade-off between the the average latency across five runs on the decontaminated DBPedia-Entities (349 queries, 1M documents) and the effectiveness of each approach.
It confirms the findings of MICE original work: MICE already stands on the same Pareto frontier as its cross-encoder counterparts \textit{without} pre-computing document representations.
In this work, we further show that when \textit{actually} using PLAID to store these document representations and measuring the retrieval latency of retrieving the top 100 passages with SPLADE, decompressing them from the PLAID index and finally scoring these passages with MICE, \micep{} is able to reach a comparable efficiency-effectiveness trade-off as state-of-the-art late-interaction approaches, like \texttt{LateOn}. 
In particular, \micep{}-130m appears as an alternative to both \texttt{LateOn-149m} or \texttt{Ettin-reranker-150m}, as it either improves significantly the latency compared to the cross-encoder and \texttt{LateOn} \textit{without} PLAID, or improves significantly the effectiveness compared to \texttt{LateOn} when allowing PLAID.  


\noindent \textbf{Storage.}  As \micep{} requires the usage of a second SPLADE index for retrieval, we further detail in \Cref{fig:pareto_mem} the memory costs for \micep{} (SPLADE index and PLAID document store), compared to those of LateOn (PLAID plays both role). Most of the difference is due to the higher precision level needed for \micep{} as discussed  in \Cref{fig:abl_nbits}, while the final size is not excessively higher.
This in turn, confirms  that \micep{} is both an efficient and  viable end-to-end retrieval architecture, positively answering \textbf{RQ1}.

\noindent \textbf{ColBERT Collapse.} It is worth noting that, while competitive when evaluated with SPLADEv3 as a first stage, our ColBERT baselines collapse when used with PLAID as first stage. We believe this is due to the non-contrastive Pearson pointwise loss: the state-of-the-art \texttt{LateOn} model, trained with InfoNCE and hard negatives \cite{sourty2026denseonlateonfullyopen}, indeed shows a significantly smaller drop when using PLAID as a first stage, suggesting that it learned to represent document vectors in a better-separated space.

\section{Conclusion}
In this work, we study and extend MICE architecture \cite{emnlpmice} from two perspectives. 
First, we successfully train ModernBERT-based MICE models whose effectiveness is on par with cross-encoders from the \texttt{Ettin-reranker} family \cite{aarsen2026ettin-reranker}, while requiring only  $\sim$ 60\% of their FLOPs. 
Second, we introduce \micep{}, an end-to-end retrieval system, where contextualized document representations are pre-computed and compressed using PLAID \cite{santhanam2022plaid}. At inference time, \micep{} only computes query representations before interaction with candidates retrieved by SPLADEv3. While \micep{} reduces  effectiveness due to the PLAID compression and increases the storage footprint, the resulting latency gains place \micep{} in a new region of the Pareto frontier, combining cross-encoder-like effectiveness with latency approaching that of ColBERT.
Future work will investigate further improving the efficiency-effectiveness trade-off of \micep{} by better compressing document representations, or enabling direct retrieval with PLAID.

\section*{Acknowledgements}
The authors acknowledge the ANR – FRANCE (French National Research Agency) for its financial support of the GUIDANCE project n°ANR-23-IAS1-0003 as well as the Chaire Multi-Modal/LLM ANR Cluster IA ANR-23-IACL-0007. This work was granted access to the HPC resources of IDRIS under the allocations 2025-A0191016944, 2024-AD011015440R1 and 2025-AD011014444R2 made by GENCI. The authors also gratefully acknowledge the support of the Centre National de la Recherche Scientifique (CNRS) through a research delegation awarded to J. Mothe.

%
%
\bibliographystyle{splncs04}
\bibliography{biblio}

\end{document}